\documentclass[superscriptaddress,prb,twocolumn,epsfig,epsf]{revtex4}
\usepackage{amsmath,amsthm,amsfonts,amscd,mathrsfs,pifont,bm}
\usepackage{eucal,graphicx,color}
\usepackage{esint}
\makeatletter
\@for\next:={int,iint,iiint,iiiint,dotsint,oint,oiint,sqint,sqiint,
ointctrclockwise,ointclockwise,varointclockwise,varointctrclockwise,
  fint,varoiint,landupint,landdownint}\do{%
    \expandafter\edef\csname\next\endcsname{%
      \noexpand\DOTSI
      \expandafter\noexpand\csname\next op\endcsname
      \noexpand\ilimits@
    }%
  }
\makeatother

\newcommand{\beq}{\begin{equation}}
\newcommand{\eeq}{\end{equation}}

\newcommand{\deln}{\delta _{m}}
\newcommand{\delnM}{\delta_{m\rm M}}

\begin{document}

\title  {Plasmonic Time Crystals}

\author {Joshua Feinberg\footnote{ https://orcid.org/0000-0002-2869-0010} }
\affiliation
       {Department of Physics and Haifa Center for Physics and Astrophysics,\\ 
University of Haifa, Haifa 3498838, Israel}

\author{David E. Fernandes}
\affiliation{University of Lisbon and Instituto de Telecomunica\c{c}\~oes, Avenida Rovisco Pais 1, Lisboa, 1049-001, Portugal}

\author{Boris Shapiro}
\affiliation{Department of Physics\\ 
Technion, Israel Institute of Technology, Haifa 32000, Israel}

\author{M\'ario G. Silveirinha}
\affiliation{University of Lisbon and Instituto de Telecomunica\c{c}\~oes, Avenida Rovisco Pais 1, Lisboa, 1049-001, Portugal}

\date   {\today}

\begin  {abstract}
We study plasmonic time crystals, an extension of dielectric-based photonic time crystals to plasmonic media.  Remarkably, we demonstrate that such systems may amplify both longitudinal and transverse modes.
In particular, we show that plasmonic time crystals support \emph{collective resonances} of longitudinal modes, which occur independently of the wave vector $k$, even in the presence of significant dissipation. These resonances originate from the coupling between the positive- and negative-frequency branches of the plasmonic dispersion relation of the unmodulated system and from the divergence of the density of states near the plasma ($\varepsilon$-near zero) frequency $\omega_p$. The strongest resonance arises at a modulation frequency $\Omega = 2 \omega_p$, corresponding to a direct interband transition. 
We demonstrate these resonances for various periodic modulation profiles and provide a generic perturbative formula for resonance widths in the weak modulation limit. 
Furthermore, we propose transparent conducting oxides as promising platforms for realizing plasmonic time crystals, as they enable significant modulation of the electron effective mass while maintaining moderate dissipation levels. 
Our findings provide new insights into leveraging time-modulated plasmonic media to enhance optical gain and control wave dynamics at the nanoscale.
\end{abstract}

\maketitle
{\it Introduction.} Temporal modulation of a material amounts to changing in time some parameters of the material. In this way, a new artificial (or {\em meta}-)material is created, with new and useful properties \cite{1,2}. In particular, if the dielectric permittivity $\varepsilon$ of a material undergoes periodic modulation, then a photonic time crystal (PTC) is produced\cite{3}. A PTC can amplify electromagnetic waves with wavenumbers $k$ lying in its $k$-band gaps\cite{4}. $k$-bands are analogous to the frequency bands in a spatial photonic crystal.

Generally, the concept of temporal modulation applies to arbitrary time dependence $\varepsilon (t)$. For instance, the behavior of an EM wave under sudden change of $\varepsilon$ at some instant $t=t_0$, has been considered already in 1958\cite{5}. The case of $\varepsilon (t)$ being a random function of time has been studied recently in \cite{3,6}. By now, temporal modulation of materials has become a broad and active field of research with many theoretical\cite{3,4,5,6,7,8,9,10,11,12,13,14,15,16,17, Solis, Solis2, Monticone, Sloan, Simon, Ptitcyn, Stanic, Holberg, Prud, Prud1, Mendoca} and some experimental\cite{18,19,20,21} works.

A few previous works have investigated some particular forms of dispersive time-crystals, focusing mainly on transverse waves \cite{Solis, Huanan, Serra2, Serra, Feng, Kim, WangX}. Let us also mention possible amplification of localized plasmons\cite{Salandrino}, as well as surface plasmons\cite{BarHillel2024}. This ongoing work on parametric amplification in various types of PTC's must be distinguished from the old work on parametric resonances and instabilities in gaseous plasmas (see e.g.\cite{Silin}). 

In this Letter, we extend the study of PTC to longitudinal excitations of an electron gas. 
In conventional PTCs, which rely on permittivity modulation in dielectric materials, both the modulation depth and frequency face stringent constraints. Specifically, both must be sufficiently large to produce observable $k$-gaps and significant amplification rates. Here, we demonstrate that \emph{plasmonic time crystals} (PLTCs) serve as exceptional platforms for generating optical gain. More precisely, we show the possibility of achieving a \emph{collective resonance} that can significantly enhance parametric gain. This collective resonance arises from the coupling of plasmons with positive and negative frequencies and is characterized by a diverging density of states. Such a resonance can occur for any excitation with a $k$-independent dispersion relation. In particular, it applies to \emph{longitudinal} plasmons, whose temporal modulation, to the best of our knowledge, has not been previously studied. 

Infinite bandgaps for transverse waves were analyzed in a recent work \cite{WangX}. Here, we show that, unlike the transverse resonant case \cite{WangX}, longitudinal waves support a collective resonance even in the presence of dissipation. Thus, for a longitudinal plasmon, a very wide (theoretically ``infinite") $k$-gap emerges naturally, leaving only the challenge of ensuring a substantial amplification rate. We also note that the temporal modulation of longitudinal optical phonons was analyzed in \cite{Huanan2} in the context of the interaction of a point charge with a time-modulated medium.

\vspace{0.2cm}
{\it Physical model.} 
We consider a conducting material whose electric response is determined by both bound and free electrons. We assume that the response of the conducting material is modulated in time via a nonlinear process driven by optical pumping. In this context, transparent conducting oxides (TCO) have recently attracted considerable attention \cite{Jaffray2022, Kinsey2015, Caspani2016, Alam2016, Zhou2020, Tirole2023, Lustig2023}, as it has been demonstrated that optical pumping can induce strong refractive index modulations in the near-zero permittivity (ENZ) regime \cite{silveirinha2006tunneling, engheta2013pursuing}. Moreover, recent studies have shown that time-interfaces as short as a single optical cycle can also be generated with such a modulation scheme \cite{Lustig2023}.

The nonlinearity of ENZ materials excited below the bandgap is governed primarily by the modulation of their absorption properties and by the modulation of the effective mass of free electrons \cite{Secondo2020}. In a plasma subjected to optical pumping, a significant fraction of conduction electrons is excited to higher energy states within the conduction band, forming a population of ``hot electrons''. Due to the nonparabolicity of the conduction band, these ``hot electrons'' may experience an effective mass $m^*$ that differs from its equilibrium value. Furthermore, the increased energy of the hot electron population can enhance the probability of lattice collisions, rendering the collision frequency $\nu$ a time-dependent parameter. 

In this Letter we concentrate on homogeneous and isotropic media. Consequently, electromagnetic excitations in our system are either longitudinal or transverse. In what follows, we shall focus mainly on longitudinal plasmons. (Transverse plasmons will be discussed mainly in comparison with longitudinal plasmons.) 

\begin{figure}[h!]
\includegraphics[scale=1.0]{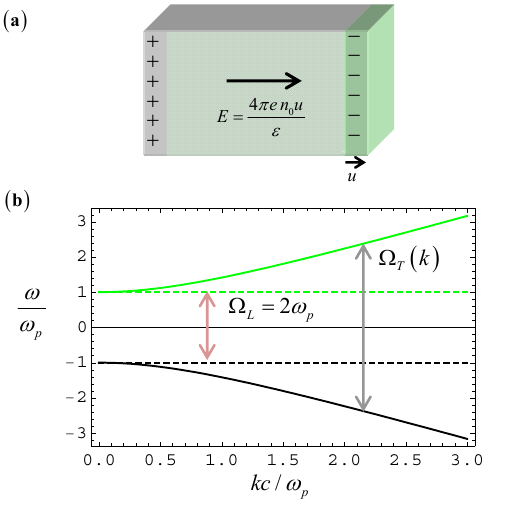}
\caption{ {\bf{a)}}
Illustration of the displacement of a region of the electron gas. The displacement generates a region of thickness $u$ with an excess of negative charge and a corresponding region of equal thickness with positive charge due to the uncompensated ionic background.
{\bf{b)}} Band structure of the static dispersive crystal, showing both positive and negative frequency bands. The arrows indicate possible interband transitions.}
\label{geomband}
\end{figure}

\vspace{0.2cm}
{\em Longitudinal Plasmons.} For this type of excitation, the magnetic field in the material is negligible, and the electron displacement away from equilibrium oscillates along the same direction as the electric field in space, as we show below. In order to derive the equation governing the dynamics of (volume) longitudinal plasmons in a material\cite{Ashcroft, Kittel}, consider some volume of the electron gas displaced by an amount $\mathbf{u}$ relative to the neutralizing ionic background, as illustrated in Fig.~\ref{geomband}a. (Note that the displaced electrons can belong to an arbitrary region within the material.) By applying Newton’s law to the displaced electrons we obtain
$
M\frac{d^2 \mathbf{u}}{dt^2} + M \nu \frac{d \mathbf{u}}{dt} = Q\mathbf{E},
$
where $M = m^*N_e$ is the total effective mass of the displaced electrons, $ Q = -e N_e$ is their total charge, and $N_e$ is the number of displaced electrons. The term proportional to $\nu \frac{d \mathbf{u}}{dt}$ accounts for dissipation due to collisions. 

The electric field $\mathbf{E}$ acting on the displaced electrons is obviously the field generated by the excess of negative charge in a layer of thickness $u$, and by the corresponding layer of uncompensated positive ionic charge of the same thickness. (see Fig.~\ref{geomband}a.) Elementary electromagnetic analysis then yields $\mathbf{E} = e\frac{4\pi n_0\mathbf{u}}{\varepsilon},$
where $n_0$ is the density of free electrons, and $\varepsilon = 1 + 4\pi \chi_e$ represents the permittivity of the background with the contribution from bound electrons described by the electric susceptibility $\chi_e$. (We adopt Gaussian units.) This relation essentially means that the total electric displacement field $\mathbf{D} = \varepsilon \mathbf{E} - 4\pi e n_0\mathbf{u} =0$ vanishes throughout the plasma medium. 

Upon substituting the relation between $\mathbf{E}$ and $\mathbf{u}$ in the equation of motion, we find that the electron displacement obeys the equation:
\begin{eqnarray}
m^* (t) \frac{d^2 \mathbf{u}}{dt^2} + m^* (t) \nu (t) \frac{d\mathbf{u}}{dt} + e^2 \frac{4\pi n_0}{\varepsilon (t)} \mathbf{u} = 0. \label{master}
\end{eqnarray}
We have incorporated the effect of optical pumping by allowing the effective mass and collision frequency to vary with time. Furthermore, for completeness, we also consider the impact of optical pumping on the susceptibility of the bound electrons, as in conventional nonlinear optics. As a result, optical pumping can also modulate the permittivity $\varepsilon$ of the background, although this effect is typically less significant than the modulation of the free electron response.

In the Supplementary Material (SM)\cite{SM} we provide a more formal derivation of \eqref{master} governing longitudinal plasmons by considering Maxwell's equations coupled with a transport equation with time-varying coefficients. 

For simplicity, in the main text, we neglect the modulation of the bound electrons permittivity, and thus set $\varepsilon$ to its unmodulated equilibrium value $\varepsilon_0$. (Modulation of $\varepsilon$ is qualitatively similar to modulation of the effective mass \cite{SM}.)  Furthermore, we also assume that $\nu$ is time-independent, treating it as a time-averaged collision frequency. The theoretical justification for this approximation is provided in the SM \cite{SM}. Under these assumptions, Eq.~\eqref{master} can be rewritten as
\begin{equation}\label{divergence1}\frac{d^2 f}{d t^2} + \omega_p^2 \left[1+\delta_m(t) \right]f + \nu\frac{d f}{d t} =0
\end{equation}
where $f$ is any component of the displacement vector $\mathbf{u}$. Here, $\omega_p$ is the plasma frequency of the unmodulated (equilibrium) system, defined as
$ \omega_p^2 = \frac{4\pi e^2 n_0}{m^*_0 \varepsilon_0}$, where $m_0^*$ is the effective mass of electrons in the unmodulated system, and where we have parametrized the modulated effective mass as
$
1/m^*  = {{\left[ {1 + \delta _m \left( t \right)} \right]} {\left/ {\vphantom {{\left[ {1 + \delta _m \left( t \right)} \right]} {m_0^* }}} \right.} {m_0^* }}$. We shall refer to the dimensionless fractional quantity $\delta_m (t)$ as the modulation depth (or modulation profile). 

Equation~\eqref{divergence1} describes a damped parametric oscillator\cite{LL}.  The first-order damping term in \eqref{divergence1} may be eliminated by redefining $f = e^{-\nu t/2} g$. The resulting equation for $g$ is 
\begin{equation}\label{divergenceg}
\frac{d^2 g}{d t^2} + \omega_p^2\left(1-\frac{\nu^2}{4\omega_p^2} + \deln(t)\right)g =0\,.
\end{equation}

Since our plasma is spatially homogeneous, it suffices to consider waves with a spatial dependence of the form $\sim e^{i{\bf k}\cdot{\bf r}}$. A distinctive feature of Eq.~\eqref{divergence1} is that it does not involve the wave number $k$. This property is unique to \emph{longitudinal} plasmons, owing to their intrinsic $k$-independent dispersion relation, $\omega = \pm\omega_p$ (see the dashed lines in Fig.~\ref{geomband}b). Such a dispersion implies that once time modulation is switched on, the function $f(t)$—along with all other relevant quantities—can undergo exponential growth simultaneously for \emph{all} $k$-modes. This collective resonance enhances the interaction between the plasma and the optical pump, leading to strong parametric amplification, as we elaborate below.

In particular, it is important to emphasize that under periodic modulation $\delta_m(t)$ with modulation frequency $\Omega$, all $k$-modes of a longitudinal plasmon excitation are amplified at the same frequency $\Omega_{\rm ampl}$. This is the collective longitudinal resonance alluded to in the Abstract. The overall scale of $\Omega_{\rm ampl}$ is essentially fixed by the two dimensionful quantities in \eqref{divergence1}, namely, $\omega_p$ and $\nu$\cite{scale}. The latter in turn are determined by the ENZ condition $\varepsilon(\omega) = 0$ (on the unmodulated permittivity), even in the presence of dissipation\cite{ENZcond}. In contrast, the related concept discussed in Ref.~\cite{WangX} relies on the condition $\varepsilon(\omega) = \infty$, which can only be achieved in the absence of dissipation.

Under periodic modulation profile $\delta_m(t)$, the two Floquet-type fundamental solutions of \eqref{divergence1} are of the form \cite{Brillouin, vdPS, Hill, SM} $f_{1,2}(t) = e^{-i\omega_{1,2} t} u_{1,2}(t)$, where $\omega_{1,2}$ are the two fundamental frequencies, and $u_{1,2}(t)$ are periodic functions with period $T=2\pi/\Omega$ \cite{SM}. 
The fundamental frequencies $\omega_{1,2} = \omega_{1,2}' + i\omega_{1,2}''$ are typically complex. Thus, if (at most) one of the imaginary parts $\omega_{1,2}''$ is positive, 
the corresponding Floquet solution will grow as function of time, rendering the system unstable.

The eigenfrequencies $\omega_{1,2}$ 
depend on the modulation frequency $\Omega$ 
and on the parameters controlling the (dimensionless) modulation profile $\deln(t)$. This parameter space is split into regions of stability and instability, separated by borderlines. In regions of stability, oscillations of $f(t)$ are bounded, while in regions of instability $f(t)$ exhibits resonant behavior, with unbounded oscillations. 
These general statements are demonstrated succinctly by the following two examples:

\begin{figure*}[t!]
\includegraphics[scale=0.8]{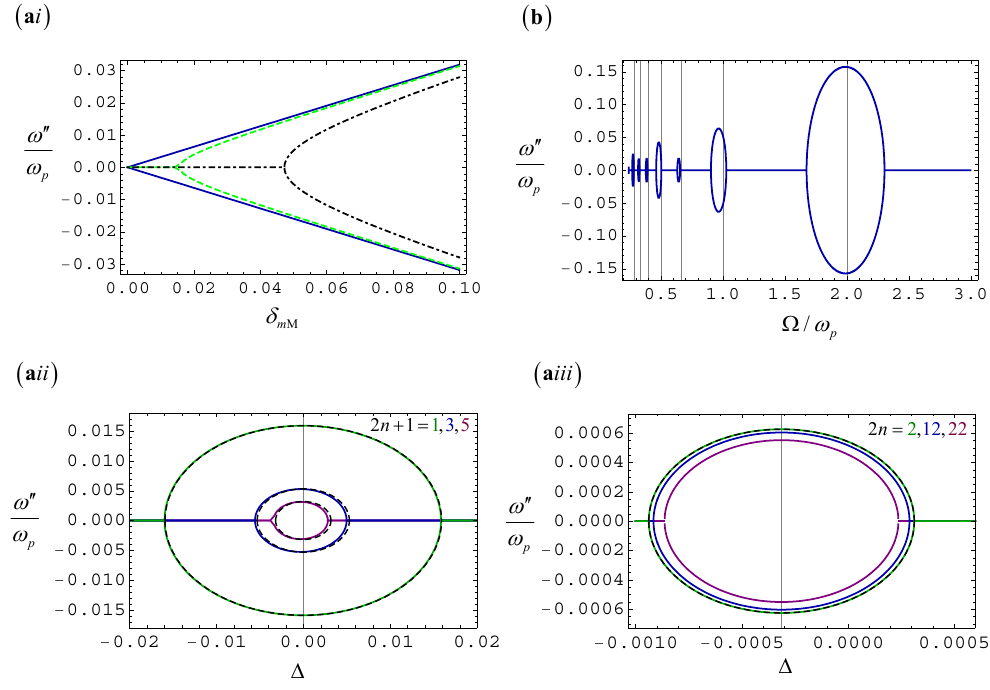}
\caption{ {\bf{ai)}} $\omega''$ as a function of the (weak) modulation strength for i) $\Omega= 2 \omega_p$ (blue solid lines), ii) $\Omega= (2 \pm 0.01) \omega_p$ (green dashed lines) and iii) $\Omega = (2 \pm 0.03) \omega_p$ (black dotted lines). {\bf{aii)}} $\omega''$ as a function of the detuning parameter $\Delta$ for fixed weak modulation amplitude $\delnM  = 0.05$ for the first three odd resonances ($2n+1=1,3,5$) in \eqref{growth1}. The dashed curves are the analytic predictions. The three resonances are centered at $\Delta=0$ and the growth rate becomes smaller as $n$ increases. {\bf{aiii)}} The same as in {\bf{bii)}} with $\delnM  = 0.05$ for the even resonances $2n=2,12,22$ in \eqref{growth3}. Note the smaller vertical scale compared to the previous case. The three displayed resonances are in good agreement with the $n$-independent prediction of \eqref{growth3}, despite the large disparity of chosen values for $n$. Evidently, these three resonances are centered around a negative value of $\Delta$ consistent with $-\delnM^2/8$ as predicted in \eqref{growth3}.
{\bf b)}
$\omega''$ as a function of the modulation frequency $\Omega$ for fixed {\em strong} modulation amplitude $\delnM  = 0.5$. The vertical grid lines mark the resonant modulation frequencies $\Omega = 2\omega_p/m$ with $m = 1,2,...,7$ for the weak modulation case.}

\label{FigLong}
\end{figure*}

\vspace{0.2cm}
(i) {\em harmonic modulation}: $\deln(t) = \delnM \cos\Omega t\,.$
In this case (and in the absence of losses) Eq.~\eqref{divergence1} reduces to the famous Mathieu equation\cite{Brillouin, vdPS, Hill, Mathieu, Ioannis}, which exhibits rich structure of stability and instability regions in the space of parameters $\Omega/\omega_p$ and $\delnM$. In particular, for $\Omega=2\omega_p$, it is well known that the amplitude of oscillations grows exponentially approximately as $\exp (\delnM \omega_pt/4)$. In our problem, this instability can be interpreted as a collective resonance (i.e., for all values of the wavenumber $k$) due to the interband transitions between the two branches $\pm\omega_p$ of the static (unmodulated) crystal (see Fig. \ref{geomband}a) \cite{Winn1999, Serra}. This instability persists also when $\Omega$ is slightly detuned away from $2\omega_p$, albeit with a smaller growth exponent\cite{LL} 
$\omega''\simeq \frac{\omega_p}{4}\sqrt{\delnM^2-16\Delta^2}$,
where $\Delta = (\Omega/2\omega_p)-1$ is the detuning parameter. 

The width of the instability region around $\Omega=2\omega_p$ is determined by reality of $\omega''$. In particular, this means that there is a minimal threshold value of $\delnM^{\rm Th} = 4|\Delta|$ required to induce instability. Similar but weaker instabilities occur also at modulation frequencies around $\Omega_n = 2\omega_p/n$ with integer $n$\cite{LL}. Finally, such exponential growth, albeit with smaller amplification rate $\omega''(\nu) \simeq \omega''(0)-\nu/2$, persists also for sufficiently small $\nu$, as should be clear from the argument leading to \eqref{divergenceg} \cite{LL}. 

In particular, it follows that for a sinusoidal-type modulation of the effective mass, the minimum modulation depth required to overcome losses (at resonance) is approximately $\delta_m \approx 2 \nu/\omega_p$. The ratio $\nu/\omega_p$ can be related to the imaginary part of the complex permittivity, $\varepsilon''(\omega)$, evaluated at the ENZ frequency via
$\frac{\nu}{\omega_p} \approx \frac{\varepsilon''_{\mathrm{ENZ}}}{\varepsilon}$,
where $\varepsilon$ is the permittivity due to bound electrons. Consequently, the required modulation depth is
$\delta_m \approx 2 \frac{\varepsilon''_{\mathrm{ENZ}}}{\varepsilon}$.
For typical TCOs, $\varepsilon \sim 3$--$4$ and $\varepsilon''_{\mathrm{ENZ}} \sim 0.2$--$0.5$ \cite{Kinsey2015, Alam2016}.
According to Ref.~\cite{Secondo2020}, for TCOs $\delta_m$ varies linearly with the pump intensity and can be as large as $0.2$. Therefore, while challenging, attaining the required modulation depth for optical gain appears feasible in these materials.

\vspace{0.2cm}
(ii) {\em periodic piecewise constant modulation}:
In this case, in its simplest form, $\deln(t)$ assumes one constant value $\delta_{m1}$ during the first part $0\leq t < \tau$ of the modulation period and another constant value $\delta_{m2}$ during its remaining part $\tau \leq t <T$\cite{Brillouin, vdPS}.
For concreteness, let us focus on the modulation 
\begin{equation}\label{piecewise}
\deln\left( t \right) = \delnM\, {\mathop{\rm sgn}} \left( {\sin \left( {\Omega t} \right)} \right)
\end{equation} 
which flips sign at the middle of the modulation period. Here, as in the case of Mathieu's equation, one obtains a rich chart of stability and instability regions in the plane of parameters $\Omega/\omega_p$ and $\delnM$\cite{Brillouin, vdPS}. In particular, for weak modulation amplitude $\delnM$ (and in the absence of dissipation $\nu=0$), we find an infinite family of resonances analogous to the aforementioned resonance of the Mathieu equation, centered at modulation frequencies $\Omega_n^{\rm odd} = 2\pi/T= 2\omega_p/(2n+1)$ with integer $n$, and with growth exponents\cite{SM}
\begin{equation}\label{growth1}
\omega'' = \omega_p\sqrt{\frac{\delnM^2}{((2n+1)\pi)^2} - \Delta^2}\,, 
\end{equation}
where $\Delta = \Omega/\Omega_n^{\rm odd} -1$ is the detuning parameter. Thus, at the center of the resonance ($\Delta = 0$), $\omega''_{\rm Res} = \omega_p \delnM/(2n+1)\pi$ is linear in $\delnM$, and is therefore of leading order in perturbation theory. Furthermore, 
similarly to the Mathieu case, detuning the modulation frequency away from the resonance requires a threshold value $\delnM^{\rm Th} = (2n+1)\pi|\Delta|$ of the modulation amplitude to induce instability with diminished $\omega''$. 

In addition to this family of `linear' resonances, we find that there is yet another family of weaker resonances centered in the vicinity of modulation frequencies $\Omega_n^{\rm even} = 2\omega_p/2n = \omega_p/n $ with integer $n$, with growth exponents \cite{SM}

\begin{equation}\label{growth3}
\omega'' = \omega_p\sqrt{\left(\frac{\delnM^2}{4}\right)^2 - \left(\Delta + \frac{\delnM^2}{8}\right)^2}
\end{equation}
where $\Delta = \Omega/\Omega_n^{\rm even}-1$ is the detuning parameter away from $\Omega_n^{\rm even}$. In contrast to \eqref{growth1}, the centers of these resonances (i.e., maximal instability) occur at modulation frequencies $\Omega_n^{\rm Res}=\Omega_n^{\rm even}(1-\delnM^2/8)$, which depend quadratically on $\delnM$. Furthermore, at $\Omega=\Omega_n^{\rm Res}$, all these resonances share an {\em $n$-independent} common growth exponent $\omega''_{\rm Res} = \omega_p \delnM^2/4$, quadratic in $\delnM$, and therefore of higher order in perturbation theory. 

The predictions of \eqref{growth1} and \eqref{growth3} are compared against numerical simulation of the piecewise constant modulation in Fig.~\ref{FigLong}ai-iii, showing excellent agreement. 

In Fig.~\ref{FigLong}b we show numerical results for the growth rates $\omega''$ of resonances arising at fixed {\em strong} modulation amplitude $\delnM=0.5$ as function of the modulation frequency $\Omega$. The general pattern of resonances at weak modulation amplitudes, summarized in \eqref{growth1} and \eqref{growth3}, is clearly preserved, up to some amount of bounded shifts in resonance frequencies. However, as can be clearly seen in Fig.~\ref{FigLong}, at strong modulation amplitude, the strength of the even resonances of \eqref{growth3} can become comparable, or even surpass the odd resonances of \eqref{growth1}. Furthermore, under strong modulation, the gain rate no longer scales linearly with $\delnM$. Hence, the weak-modulation estimate of the gain typically underestimates the actual gain, potentially facilitating a practical realization.

In the SM we give a simple perturbative derivation of the growth rate $\omega''$ of `linear resonances' (at zero detuning $\Delta =0$ and in the absence of losses $\nu=0$) for an arbitrary weak periodic modulation profile $\deln(t)$ in \eqref{divergence1} \cite{SM}. Our analysis confirms that the instabilities generally occur when the modulation frequency $\Omega$ is commensurate with the gap width $2\omega_p$ of the unmodulated longitudinal plasmons, such that $2\omega_p/\Omega = n$, with integer $n$. Specifically, the growth rate of the $n$th resonance is governed by the $n$-th Fourier coefficient of $\deln(t)  = \sum_{l=1}^\infty \left(c_l \exp (il\Omega t) + c_l^* \exp (-il\Omega t)\right)$, such that
\begin{equation}\label{growth4}
    \omega''_n = \frac{\omega_p |c_n|}{2}\,.
\end{equation}
This result is linear in $c_n$, and therefore in $\deln$, because it arises from first order perturbation theory. The above formula includes as particular cases the harmonic (Mathieu) modulation and the piecewise constant modulation, discussed previously \cite{SM}.
Moreover, in \cite{SM}, we demonstrate that for a given peak modulation depth $\delnM$ , the optimal modulation profile that maximizes optical gain corresponds to a piecewise constant modulation.


\vspace{0.2cm}
{\em Transverse Waves.}~~We shall now briefly consider temporal modulation of transverse waves\cite{Solis, Huanan, Serra2, Serra, Feng, Kim}, mainly in order to compare it against that of longitudinal plasmons. We shall assume in this section that the material dissipation is negligible ($\nu=0$). 
For transverse waves, the oscillations of free electrons occur along a direction perpendicular to the spatial field gradients, i.e., perpendicular to the wave vector ${\bf k}$. 
A detailed analysis presented in the SI shows that the time dynamics of transverse waves is governed by \cite{SM}:
\begin{equation}\label{divergence3}
\frac{{d^2 {\bf{D}}_{\rm{b}} }}{{d t^2 }} + \omega _T^2 \left[ {1 + \frac{{\omega _p^2 }}{{\omega _T^2 }}\delta _m \left( t \right)} \right]{\bf{D}}_{\rm{b}}  = 0,
\end{equation}
where $\omega_T^2 = \frac{c^2 k^2}{\varepsilon_0}+ \omega_p^2$ is ($k$ dependent) transverse wave dispersion, and ${\bf{D}}_{\rm{b}} = \varepsilon {\bf{E}}$, where ${\bf{E}}$ is the electric field. 
The plane wave spatial variation $\sim e^{i{\bf k}\cdot{\bf r}}$ of the fields is implicit. Furthermore, analogous to the case of longitudinal plasmons, in the following we suppose that $\varepsilon$ remains time-independent,

Equation \eqref{divergence3} is qualitatively similar in form to \eqref{divergence1} (with $\nu=0$). It coincides with the latter at $k=0$, and
therefore describes a parametric oscillator, as in the case of longitudinal plasmons. Evidently, we can translate the entire previous discussion of resonances in modulated longitudinal plasmons to transverse plasmons, simply by replacing $\omega_p$ in the appropriate places by $\omega_T$ and by rescaling the modulation amplitude $\deln(t)$ by a factor $\omega_p^2/\omega_T^2$. Thus, compared to the longitudinal case, the modulation amplitude is effectively reduced and the resonance frequency of the system is increased. 
Stability regions in the parameter space pertaining to modulated longitudinal plasmons  become $k$-bands of the transverse PLTC, while regions of instability correspond to gaps.

Importantly, the resonances associated with Eq.~\eqref{divergence3} arise independently for each wavenumber $k$, in contrast to the collective, $k$-independent nature of resonances in modulated longitudinal plasmons. As a result, the resonances associated with transverse waves are weaker, both due to the reduced modulation depth and because they remain isolated from one another.

For example, in the case of piecewise constant modulation \eqref{piecewise}, we see from \eqref{growth1} pertaining to longitudinal plasmons, that there will be resonances of the transverse plasmon at modulation frequency $\Omega_{n,T}^{odd} = 2\omega_T/(2n+1)$ with integer $n$ and growth exponents $\omega'' =\omega_T \sqrt { (\omega_p^2\delnM/(2n+1)\pi\omega_T^2)^2- \Delta^2}$, in which $\Delta = \Omega/\Omega_{n,T}^{\rm odd} - 1$ is the detuning parameter. Analogous results exist obviously also for the weaker longitudinal resonances \eqref{growth3}. A numerical example is reported in the SI \cite{SM}.

{\em Conclusions.}~~~
In this Letter, we have carried out a comprehensive study of \emph{plasmonic time crystals}, demonstrating that these platforms support both longitudinal and transverse modes. We have shown that under periodic time modulation, these systems function as parametric oscillators, and we have analyzed their stability properties across a broad range of parameters. 

Notably, the resonance of longitudinal modes is independent of the wave vector $\mathbf{k}$, even in presence of significant dissipation, enabling a \emph{collective resonance} that can be leveraged to achieve optical amplification. The strongest parametric instabilities arise from interband transitions at modulation frequency $\Omega = 2 \omega_p$. In particular, we have proposed that TCOs are well-suited for realizing plasmonic time crystals, as they allow for significant modulation of the electron effective mass while maintaining a moderate dissipation level \cite{Jaffray2022, Kinsey2015, Caspani2016, Alam2016, Zhou2020, Tirole2023, Lustig2023}.

In contrast, transverse resonances depend on the wave vector $\mathbf{k}$ of propagating waves and remain isolated from one another, limiting their capacity for collective amplification. Our findings pave the way for the exploration of plasmonic time crystals as a platform for tunable optical amplification in nanophotonics.

\begin{acknowledgments}
BS is grateful for the hospitality extended to him during his visit to the University of Lisbon in October 2022, where this
project was initiated. BS is also indebted to Remi Carminati for useful discussions at the early stage of this work. JF is supported in part by grant No.2022158 from the United States - Israel Binational Science Foundation (BSF), Jerusalem, Israel. D.E.F. acknowledges financial support by IT-Lisbon and FCT under a research contract Ref. CEECINST/00058/2021/CP2816/CT0003 and DOI identifier https://doi.org/10.54499/CEECINST/\newline 00058/2021/CP2816/CT0003.
MS is partially supported by the IET, by the Simons Foundation under the award 733700 (Simons Collaboration in Mathematics and Physics, Harnessing Universal Symmetry Concepts for Extreme Wave Phenomena), by Funda\c{c}\~ao para a Ci\^encia e a Tecnologia and Instituto de Telecomunica\c{c}\~oes under projects UIDB/50008/2020 and DOI identifier https://doi.org/10.54499/UIDB/50008/2020 and 2022.06797.PTDC. 
\end{acknowledgments}


\begin{references}
\bibitem{scale} That is, up to a dimensionless coefficient which may depend in general on other dimensionless parameters in \eqref{divergence1}, such as the amplitude of the modulation depth $\delta_m(t)$.
\bibitem{ENZcond} 
The simplest ENZ material is perhaps ordinary dissipationless electron plasma with $\varepsilon(\omega) =1-(\omega_p/\omega)^2$, with $\omega\approx \omega_p$.
\end{references}
\end{document}


\title  {Supplementary Material for `Plasmonic Time Crystals'}

\author {Joshua Feinberg\footnote{ https://orcid.org/0000-0002-2869-0010} }
\affiliation
       {Department of Physics and Haifa Center for Physics and Astrophysics,\\ 
University of Haifa, Haifa 3498838, Israel}

\author{David E. Fernandes}
\affiliation{University of Lisbon and Instituto de Telecomunica\c{c}\~oes, Avenida Rovisco Pais 1, Lisboa, 1049-001, Portugal}

\author{Boris Shapiro}
\affiliation{Department of Physics\\ 
Technion, Israel Institute of Technology, Haifa 32000, Israel}

\author{M\'ario G. Silveirinha}
\affiliation{University of Lisbon and Instituto de Telecomunica\c{c}\~oes, Avenida Rovisco Pais 1, Lisboa, 1049-001, Portugal}

\date   {\today}

\maketitle

\section{Electromagnetic Model for the Time Modulated Plasma} 

\subsection{Physical mechanisms} 

We consider a conducting material with both free and bound electrons. As usual, the microscopic currents in the material consist of two components: the polarization current, arising from bound electrons, and the conduction current, due to free electrons. Here, we assume that the response of the conducting material is modulated in time via a nonlinear process based on optical pumping. Specifically, we explore two possible mechanisms for this modulation.

The first mechanism involves time-dependent modulation of the susceptibility $\chi_e(t)$ of bound electrons, as in conventional nonlinear optics. This is typically a weak effect.

The second mechanism results from the effect of the optical pump on the free electrons. In a plasma subjected to optical pumping, a significant fraction of the conduction electrons is excited to higher energy states within the conduction band, thereby forming a population of ``hot electrons'' (see Refs. \cite{Kinsey2015, Alam2016, Lustig2023}). Optical pumping creates a non-equilibrium state and can induce time variations in the effective mass $m^*$ experienced by the hot electrons due to the non-parabolicity of the conduction band. Additionally, the increased energy of the ``hot electron'' population may enhance the probability of lattice collisions, causing electrons to lose momentum more rapidly under optical pumping. This accelerated momentum relaxation renders the collision frequency, $\nu$, a time-dependent quantity. These two effects have been suggested as the key factors controlling the nonlinearity of $\varepsilon$-near zero (ENZ) materials when excited below the bandgap \cite{Secondo2020}.

\subsection{Dynamical equations} 

The electrodynamics of the system under consideration is governed by Maxwell's equations
\begin{subequations} \label{Max}
\begin{eqnarray}  
 \nabla  \times {\bf{E}} &=&  - \frac{1}{c} \partial _t {\bf{B}} \label{Max_a} \\ 
 \nabla  \times {\bf{B}} &=& \frac{1}{c} \partial _t
 \underbrace {\left[\varepsilon \left( t \right){\bf{E}}\right]}_{\bf{D_{\rm b}}}  
+ \frac{4 \pi}{c} \partial _t {\bf{P}_{\rm c}}  \label{Max_b}
\end{eqnarray}
\end{subequations}
Here, $\bf{D_{\rm b}}$ is the electric displacement vector accounting exclusively for the effect of bound electrons, and ${\bf{J}}_c =\partial_t {{\bf{P}}_{\rm c}}$ is the conduction current expressed in terms of the polarization vector $\mathbf{P}_c$ due to the cloud of free electrons. As discussed above, the susceptibility $\chi_e$ of the bound electrons may be time-dependent, thus giving rise to a time dependent permittivity $\varepsilon \left( t \right) = 4\pi \chi_e  + 1$. (We adopt Gaussian units.)

The 
polarization vector ${\bf{P}}_{\rm c}$
is governed by a standard transport equation, but with time dependent coefficients:
\begin{eqnarray}  
m^*(t)\partial _t^2 {\bf{P}}_{\rm c} + m^*(t) \nu\left( t \right) \partial _t {\bf{P}}_{\rm c} = n_0 e^2{\bf{E}}.  \label{transport}
\end{eqnarray}
Here, $-e$ is the electron charge, $n_0$ is the electron concentration, $m^*(t)$ is the time-dependent electron effective mass and $\nu\left( t \right)$ is the time-dependent collision frequency.

The system of equations \eqref{Max}-\eqref{transport} admits two types of solutions.

\subsubsection{Longitudinal solutions} 
For a longitudinal wave, we require that:
\begin{eqnarray}  
{\bf{B}} = 0, \quad \nabla  \times {\bf{P}_{\rm c}} = 0.
\end{eqnarray}
The condition $\nabla  \times {\bf{P}_{\rm c}} = 0$ ensures that the oscillations of the free-electrons are longitudinal. (For example the polarization can be a function of the type $
{\bf{P}_{\rm c}} = P_c\left( {x,t} \right){\bf{\hat x}}$.) Then, clearly 
\begin{eqnarray}\label{displacement1}  
{\bf{E}} =  - \frac{{4 \pi}}{{ \varepsilon \left( t \right)}}{\bf{P}_{\rm c}}
\end{eqnarray}
is a solution of Amp\`ere’s law \eqref{Max_b}. Faraday’s law [Eq. \eqref{Max_a}] is automatically satisfied due to $\nabla  \times {\bf{P}_{\rm c}} = 0$. 

The condition \eqref{displacement1} is equivalent to the statement that the total electric displacement vector\cite{LLX} 
\begin{equation}\label{displacement}
\mathbf{D} = \varepsilon(t)\mathbf{E} + 4\pi \mathbf{P}_c = 0
\end{equation}
vanishes throught the plasma medium. This must be so because $\mathbf{D}$ is subjected simultaneously to both conditions $\nabla\cdot\mathbf{D}=0$ (by definition, also for our time-modulated system), and $\nabla\times\mathbf{D}=0$ (longitudinality). 

By substituting \eqref{displacement1} in the transport equation, we obtain the ordinary differential equation
\begin{eqnarray}  
\frac{d^2 {\bf{P}}_{\rm{c}}}{dt^2}  + \nu\left( t \right)\frac{d{\bf{P}}_{\rm{c}}}{dt}  + \omega _{{p}}^{\rm{2}} \left( t \right){\bf{P}}_{\rm{c}}  = 0, \quad \omega _{{p}}^{\rm{2}} \left( t \right) = \frac{{4\pi n_0 e^2 }}{{\varepsilon \left( t \right)m^*(t)}}. \nonumber \\ \label{longfinal}
\end{eqnarray}
This is the result used in the main text with ${\bf{P}}_{\rm c}=-e n_0 {\bf u}$. In the above, $\omega_{p}(t)$ denotes the instantaneous bulk plasmon frequency. In the absence of time modulation, it corresponds to the frequency at which the total permittivity vanishes\cite{Kittel}.

\subsubsection{Transverse solutions} 
For a transverse wave, we impose the condition:
\begin{eqnarray}  
\nabla  \cdot {\bf{P}}_{\rm{c}} = 0.
\end{eqnarray}
This ensures that the free electrons execute transverse oscillations. Then, from Amp\`ere’s law [Eq. \eqref{Max_b}],
\begin{eqnarray}\label{Db}  
\nabla  \cdot {\bf{D}_{\rm b}} = \nabla  \cdot \left( {\varepsilon(t) {\bf{E}}} \right) = 0.
\end{eqnarray}
By combining the two Maxwell’s equations, we obtain:
\begin{eqnarray}  
 \nabla  \times \nabla  \times {\bf{E}} &=&  - \frac{1}{{c^2 }}\partial _t^2 {\bf{D}}_{\rm{b}}  - \frac{{4\pi }}{{c^2 }}\partial _t^2 {\bf{P}}_{\rm{c}}  \nonumber \\ 
  &=&  - \frac{1}{{c^2 }}\partial _t^2 {\bf{D}}_{\rm{b}}  - \frac{{4\pi }}{{c^2 }}\left[ { - \nu(t) \partial _t {\bf{P}}_{\rm{c}}  + \frac{{n_0 e^2 }}{m^*(t)}{\bf{E}}} \right]. \nonumber \\ \label{transverseW}
\end{eqnarray}
We used the transport equation [Eq. \eqref{transport}] in the last identity. In the following, we ignore the effect of collisions on the dynamics of the transverse waves. This simplification will allows us to obtain a master equation for the transverse waves involving only ${\bf D}_{\rm b}$. (When $\nu$ is nontrivial, the different dynamical fields cannot be decoupled, and a first-order vector equation in time provides the most effective formulation. We do not analyze this case here.)

In addition, we suppose that the permittivity $
\varepsilon  = \varepsilon \left( t \right)$ is space independent in the region of interest, so that from \eqref{Db} we simply obtain that $
\nabla  \cdot {\bf{E}} = 0$. In this case, Eq. \eqref{transverseW} simplifies to:
\begin{eqnarray}  
\nabla ^2 {\bf{E}} - \frac{1}{{c^2 }}\partial _t^2 {\bf{D}}_{\rm{b}}  - \frac{1}{{c^2 }}\omega _{{p}}^2 \left( t \right){\bf{D}}_{\rm{b}}  = 0.
\end{eqnarray}  
Thus, in the absence of dissipation, transverse waves are governed by
\begin{eqnarray}  
\partial _t^2 {\bf{D}}_{\rm{b}}  + \omega _{{p}}^2 \left( t \right){\bf{D}}_{\rm{b}}  - \frac{{c^2 }}{{\varepsilon \left( t \right)}}\nabla ^2 {\bf{D}}_{\rm{b}}  = 0. \label{transvfinal}
\end{eqnarray}  
Note that due to isotropy of the problem, the three components of ${\bf{D}}_{\rm{b}}$ obey the same equation, which we can treat as a scalar problem.

Considering spatial variation of the form \( \exp (i \bf{k \cdot r}) \), so that \( \nabla  := i{\bf{k}} \), we see that transverse plasmons obey the ordinary differential equation
\begin{eqnarray}  
\frac{d^2{\bf{D}}_{\rm{b}}}{dt^2}  + \omega _{{ p},k}^2 \left( t \right) {\bf{D}}_{\rm{b}}  = 0, \label{transvfinal2} 
\end{eqnarray}  
where 
\begin{eqnarray}\label{ompk}  
\omega _{p,k}^2 \left( t \right) = \omega _p^2 \left( t \right) + \frac{{c^2 k^2 }}{{\varepsilon _0 }}\left[ {1 + \delta _\varepsilon  \left( t \right)} \right].  
\end{eqnarray}

\subsubsection{Relation between the parametric oscillators corresponding to the transverse and longitudinal waves in the absence of dissipation} 
\label{Sec_RelationLongTrans}

The two ordinary differential equations \eqref{longfinal} and \eqref{transvfinal2} describe parametric oscillators\cite{LL}. Upon setting $\nu(t)=0$ in \eqref{longfinal}, one can translate easily between these two oscillators as we now explain. 

As in the main text, we parametrize the effective mass by
\begin{eqnarray}  
m^* \left( t \right) = \frac{m^*_0}{{1 + \delta_{m} \left( t \right)}},
\end{eqnarray}  
where $m^*_0$ is the effective mass in the equilibrium state and $\delta_m(t)$ is the dimensionless modulation profile. Similarly, we write the permittivity due to the effect of the bound electrons as
\begin{eqnarray}  
\varepsilon \left( t \right) = \frac{\varepsilon_0}{{1 + \delta_{\varepsilon} \left( t \right)}},
\end{eqnarray}  
where $\varepsilon_0$ is the permittivity in the absence of modulation, and $ \delta_{\varepsilon} \left( t \right)$ is the corresponding fractional dimensionless modulation strength. Consequently, the instantaneous volume plasmon resonance frequency can be expressed as:
\begin{eqnarray}\label{ompt}  
\omega _{p}^2 \left( t \right) \approx \omega _{ p}^2 \left[ {1 + \delta_{m} \left( t \right)+\delta_{\varepsilon} \left( t \right)} \right],
\end{eqnarray}
where $\omega_p^2 = 4\pi n_0 e^2/\varepsilon_0 m_0^*$ is the square of the plasma frequency of the unmodulated system. In this paper we concentrate on weak modulation effects, and therefore only retained in \eqref{ompt} terms linear in the modulation strength. 

In the absence of dissipative effects (\( \nu \approx 0 \)), the dynamics of longitudinal plasmons is governed by (see \eqref{longfinal}):
\begin{eqnarray}  
\frac{d^2{\bf{P}}_{\rm{c}}}{dt^2}  + \omega _{{p}}^{2} (t) {\bf{P}}_{\rm{c}}  = 0,  \label{longfinal2}
\end{eqnarray}  
while that of transverse plasmons is governed by \eqref{transvfinal2}.

Let us compare \eqref{ompk} and \eqref{ompt}. For weak modulation we clearly have $\omega _{p,k}^2 (t) \geq \omega _p^2(t)>0$. 

Let us further assume that both $\delta_m(t)$ and $\delta_\varepsilon(t)$ are periodic with a common modulation frequency $\Omega$, so that both \eqref{transvfinal2} and \eqref{longfinal2} have Floquet gaps in which the corresponding systems become unstable and amplification occurs. Under such conditions, if all $k$-longitudinal plasmon modes are amplified at a certain modulation frequency $\Omega_{\rm long}$, whose scale must be proportional to $\omega_p$, the corresponding transverse system at wave vector $\mathbf{k}$ will amplify at a `blue-shifted' modulation frequency $\Omega_{\rm trans} = \sqrt{1 + c^2k^2/\omega_p^2\varepsilon_0}\,\Omega_{\rm long}$. Furthermore, by rewriting \eqref{ompk} as 
\begin{eqnarray}  \label{wpk}
\omega _{p,k}^2 \left( t \right) = \left( {\omega _p^2  + \frac{{c^2 k^2 }}{{\varepsilon _0 }}} \right)\left[ {1 + \delta _\varepsilon  \left( t \right) + \frac{{\omega _p^2 }}{{\omega _p^2  + \frac{{c^2 k^2 }}{{\varepsilon _0 }}}}\delta _m \left( t \right)} \right]. \nonumber \\
\end{eqnarray}  
it is evident that the modulation strength for the transverse plasmons (2nd and 3rd parcels of the term in rectangular brackets) is typically weaker, leading to weaker instablity growth within each modulation cycle.

\subsection{Discussion of the qualitative effect of the different modulations} 

The time modulation of $\omega_p$, described by the parameters $\delta_{\varepsilon}(t)$ and $\delta_{m}(t)$, always contributes to generate optical gain. This result follows from the analysis in Sect. \ref{Sec_perturbation} below.

On the other hand, as intuitively expected, the modulation of the collision frequency does not generate gain. This can be easily seen in the case where $\omega_p(t)$ is time independent. Indeed, multiplying both sides of Eq.~\eqref{longfinal} by $
\partial _t {\bf{P}}_{\rm{c}}$, one easily finds that

\begin{eqnarray}
\frac{\partial }{{\partial t}}\left[ {\frac{1}{2}\partial _t {\bf{P}}_{\rm{c}}  \cdot \partial _t {\bf{P}}_{\rm{c}}  + \frac{1}{2}\omega _p^2 {\bf{P}}_{\rm{c}}  \cdot {\bf{P}}_{\rm{c}} } \right] =  - \nu \left( t \right)\partial _t {\bf{P}}_{\rm{c}}  \cdot \partial _t {\bf{P}}_{\rm{c}}. \nonumber \\
\end{eqnarray}

As $\nu(t)>0$ the right-hand side of the equation is strictly negative. Thus, the positive quantity inside rectangular brackets is a strictly decreasing function of time, implying that the energy stored in the longitudinal plasmons is always dissipated when 
$\delta_{\varepsilon}(t)=\delta_{m}(t)=0$.

\subsection{Other modulation schemes} 

In principle, using an electric pump, it is also possible to modulate the electron density $n$. For example, this could be done using a sufficiently thin, grounded semiconductor slab, positioned beneath a gate plate. Together, the semiconductor and gate plate form the plates of a capacitor. By applying a time-varying voltage across this capacitor, we can modulate the charge within the semiconductor slab. Charge may be injected or extracted by grounding the semiconductor.  

The relevant dynamical equations could be derived by linearizing the transport and continuity equations coupled to Maxwell's equations. However, this type of modulation involves charge transport, which is usually too slow compared to the effect of optical pumping, and therefore, this possibility will not be further discussed here.

\section{Summary of relevant results from Floquet Theory}

\subsection{Structure of the Bloch solutions}
Let us consider a scalar version of Eq.~\eqref{longfinal}, and denote $f$ as the relevant component of the polarization vector $\bf P_{\rm c}$. The function $f$ satisfies
\begin{equation}\label{divergence1}
\frac{d^2 f}{dt^2} + \omega_p^2(t)f + \nu(t) \frac{df}{dt} =0.
\end{equation}
The first-order damping term may be eliminated by redefining 
\begin{equation} \label{defg}
f\left( t \right) = g\left( t \right)e^{ - \int\limits_{}^t {\frac{{\nu \left( \tau  \right)}}{2}d\tau } }. 
\end{equation}
In fact, it is straightforward to show that:
\begin{eqnarray}\label{divergenceg}
\frac{{d^2 g}}{{dt^2 }} + \omega _{\rm{p}}^2 \left[1 + \deln \left( t \right) \right]g = 0, 
\end{eqnarray}
with
\begin{eqnarray} \label{thetadef}
\deln \left( t \right) = \delta _m  \left( t \right)+\delta _\varepsilon  \left( t \right) - \frac{{\nu ^2 \left( t \right)}}{{4\omega _{{p}}^2 }} - \frac{{\dot\nu\left( t \right)}}{{2\omega _{{p}}^2 }}.
\end{eqnarray}

Assuming that $\delta_m(t), \delta_\varepsilon(t),\nu(t)$, and therefore $\delta_{\rm mod}(t)$, are all periodic functions oscillating at a common modulation frequency $\Omega$, Eqs.\eqref{divergence1} and \eqref{divergenceg} are subjected to Floquet theory.  

The two Floquet-type fundamental solutions of \eqref{divergenceg} are of the form \cite{Brillouin, vdPS, Hill} $g_{1,2}(t) = e^{-i \tilde\omega_{1,2} t} u_{1,2}(t)$, where $\tilde \omega_{1,2}$ are the two fundamental frequencies, and $u_{1,2}(t)$ are periodic functions with period $T=2\pi/\Omega$. 

From Eq. \eqref{defg}, it follows that $f_{1,2}(t +T) = e^{-i\omega_{1,2}T}f_{1,2}(t) = \Lambda_{1,2} f_{1,2}(t)$, where $\omega_{1,2}= \tilde \omega_{1,2} -i\nu _{{\rm{av}}} /2$, where $\nu _{{\rm{av}}}$ is the time average of $\nu(t)$ over one period $T$. In particular, $\Lambda_{1,2} = e^{-i\omega_{1,2}T}$ are the eigenvalues of the temporal transfer matrix $\M{T}$, which acts on the two dimensional space of solutions of \eqref{divergence1} and propagates them through one period of time. By studying the Wronskian $W(g_1,g_2)$ one can show that $
e^{ - i\left( {\tilde \omega _1  + \tilde \omega _2 } \right)T}  = 1$. Thus, it follows that $
\Lambda _1 \Lambda _2  = e^{ - \nu _{{\rm{av}}} T} 
$, or equivalently $\omega_1 + \omega_2 = -i\nu _{{\rm{av}}} + n\Omega$ with integer $n$. 

The coefficient functions in \eqref{divergence1} are all real, so that if $f$ is a solution, so is $f^*$. Thus, there are two possibilities: either (i) $f_2(t) = f_1^*(t)$, or (ii) $f_1(t)$ and $f_2(t)$ are both real. Case (i) corresponds to {\em stable oscillations}, for which $\Lambda_2 = \Lambda_1^*$, so that $|\Lambda_1|^2 = e^{-\nu_{\rm av} T} \leq 1$. Therefore, both $f_{1,2}(t)$ are bounded and the system is stable. In this case $e^{i(\omega_1^* + \omega_2)T} = 1$ so that we can always choose $\omega_{1,2} = \pm \omega' -i\nu_{\rm av}/2$. In contrast to case (i), case (ii) typically corresponds to {\em unstable oscillations}, where both $\Lambda_1$ and  $\Lambda_2$ are real and of the same sign. For $\nu =0$ and $\deln \ne 0$, one of them must be larger than $1$ in absolute value. In this case, one of the solutions becomes unbounded at large times, rendering the system unstable. By continuity this instability should also persist after $\nu$ is turned on, for small enough values of $\nu$.
If  both $\Lambda_1, \Lambda_2 >0$ then $\omega_1 = i\omega''$ and $\omega_2 = -i(\omega''+ \nu_{\rm av})$ are pure imaginary. Alternatively, if both $\Lambda_{1,2}$ are negative, then $e^{-i\omega_{1,2}T} = \Lambda_{1,2} = |\Lambda_{1,2}| e^{\pm i\pi}$ so that ${\rm Re}\,\omega_{1,2} = \omega_{1,2}' = \pm \pi/T = \pm \Omega/2$ (as is evident for example from the numerical results plotted in Fig.~\ref{Fig-Dissipation} in Sect \ref{Sec_piecewise}.)

\subsection{Propagator}
\label{Sec_propagator}

It is useful to adopt a ``Hamiltonian" approach and rewrite \eqref{divergenceg} as a system of two coupled first order equations
\begin{equation}\label{hamiltonian}
\frac{d}{dt}\left(\!\!\begin{array}{c}\psi_1\\{}\\ \psi_2\end{array}\right) = 
\left(\!\!\begin{array}{cc} 0 & \omega_p\\{}\\ -\omega_p\left[1+\deln(t)\right] & 0\end{array}\right) \left(\!\!\begin{array}{c}\psi_1\\{}\\ \psi_2\end{array}\right)
\end{equation}
with $\psi_1=\omega_p g(t)$ and $\psi_2=\dot g(t)$ (thus rendering $\psi_1$ and $\psi_2$ the same physical dimension). Equation \eqref{hamiltonian} can be integrated formally by applying the time ordered exponent $\M{U}(t,t_0) =  {\cal T}\exp \left(\int\limits_{t_0}^t \M{L}(t')dt' \right)$ of the matrix $\M{L}(t)$ (the ``Liouvillian") on the right hand side of \eqref{hamiltonian} to the vector of initial conditions at $t=t_0$. An example is provided in section \ref{Sec_piecewise}.


\section{Parametric Resonances at Weak Modulation - The General Case} 
\label{Sec_perturbation}

\subsection{Gain rate} 
We shall now offer a very simple demonstration of the existence of parametric resonances for generic periodic modulation at frequencies $\Omega = 2\omega_p/n$ with integer $n$, and derive the corresponding growth exponents at these resonances (i.e., at zero detuning $\Delta$) to leading order in the small modulation amplitude $\delta_{\rm mod}(t)$.
For simplicity, we limit the discussion to the non-dissipative case $\nu = 0$. To this end, we consider the system of equations \eqref{hamiltonian}, 
where now $\psi_1=\omega_p f(t)$ and $\psi_2 = \dot f(t)$. For simplicity, we suppose that the weak modulation $\deln(t)$ (see Eq.~\ref{thetadef}) oscillates equally between positive and negative values, so that its zeroth Fourier component $c_0 = \int\limits_0^T \deln(t) dt/T = 0$ (i.e., its mean value) vanishes. The general case can be reduced to this one by redefining $\omega_{p}$.

In the absence of modulation ($\deln=0$), Eq.~\eqref{hamiltonian} is identical in form to the Schr\"odinger equation for a spin-$1/2$ (with magnetic moment normalized to $2$) precessing in a constant magnetic field  $\M{B} = -\omega_p \hat {\bf y}$, whose eigen-solutions are
\begin{equation}\label{spinor}
    \Psi_{\pm}(t)  = \frac{1}{\sqrt{2}}\left(\begin{array}{c} 1\\{}\\ \pm i\end{array}\right) e^{\pm i\omega_p t}
\end{equation}
and with corresponding eigenvalues $\mp \omega_p$. Now we turn the modulation on, and in the spirit of the discussion in \cite{LL} of parametric resonances in the Mathieu equation, seek a solution of \eqref{hamiltonian} in the form  $\Psi(t) = (\psi_1,\psi_2)^T = a(t)  \Psi_+(t) + b(t) \Psi_-(t)$, with {\em slowly varying} amplitudes $a(t)$ and $b(t)$, which serve as amplitude envelopes to the harmonically oscillating factors. (This is reminiscent of transforming to the interaction picture in Quantum Mechanics.) Thus, if $f(t) = \psi_1(t)/\omega_p$ is the Floquet eigensolution of the equation with eigenvalue $\Lambda = e^{-i(\omega' + i\omega'')T}$, then the growth (or decay) coefficient $\omega''$ should be encoded in the exponential envelope of $a(t)$ and $b(t)$, while the oscillatory parts of the latter should combine with the phase factors $e^{\pm i\omega_p t}$ in \eqref{spinor} to produce the real part $\omega'$ of the Floquet frequency.  By substituting this form of $\Psi(t)$ in \eqref{hamiltonian} and utilizing orthonormality of the eigenspinors \eqref{spinor} we obtain the equation for $a(t)$ and $b(t)$ as
\begin{equation}\label{hamiltonian2}
\frac{d}{dt}\left(\!\!\begin{array}{c} a(t)\\{}\\ b(t)\end{array}\right) = 
i\frac{\omega_p\deln(t)}{2}\left(\!\!\begin{array}{cc} 1 & e^{-2i\omega_p t}\\{}\\ -e^{2i\omega_p t} & -1\end{array}\right) \left(\!\!\begin{array}{c} a(t)\\{}\\ b(t)\end{array}\right)\,.
\end{equation}
This equation is exact. In what follows we shall assume the modulation $\deln(t)$ is very weak, and contend ourselves with solving \eqref{hamiltonian2} perturbatively to leading order in $\deln(t)$.

By assumption, $\deln(t)$ oscillates evenly between positive and negative values, and its oscillations are in general incommensurate with those of the phases $e^{\pm 2i\omega_p t}$. In our perturbative solution of \eqref{hamiltonian2}, we have to integrate its right-hand side over a period of time $t$ (starting at some initial time $t_0$). Let us assume  $t$ contains many modulation cycles. For generic modulation $\deln(t)$ it is plausible to expect significant cancellations (or `destructive interference') in this integral. Therefore, this integral should be dominated by the least oscillating terms in the quantities $\deln(t)\exp (\pm 2i\omega_pt)$ in the integrand, which must be a combination of an appropriate Fourier mode of $\deln(t)$ and the phases $\exp (\pm 2i\omega_pt)$.  Thus, we substitute the Fourier decomposition $\deln(t)  = \sum_{l=1}^\infty \left(c_l \exp (il\Omega t) + c_l^* \exp (-il\Omega t)\right) $ in \eqref{hamiltonian2} and average it over one modulation period. In this procedure we encounter integrals of the form $(1/T)\int\limits_0^T dt \exp [\pm i(l\Omega -2\omega_p)t]$, whose phase is minimized for the pair of Fourier modes corresponding to $l =  [2\omega_p/\Omega]$ (where $[x]$ is the integral part of the real number $x$). In particular, if $\Omega$ and $2\omega_p$ are commensurate such that $\Omega = 2\omega_p/n$ with integer $n$, this minimal phase, occurring for $l=n$ will be exactly null, and the dominant part in \eqref{hamiltonian2}
will be 
\begin{equation}\label{hamiltonian3}
\frac{d}{dt}\left(\!\!\begin{array}{c} a(t)\\{}\\ b(t)\end{array}\right) = 
\frac{\omega_p}{2}\left(\!\!\begin{array}{cc} 0 & ic_n\\{}\\ -ic^*_n & 0\end{array}\right) \left(\!\!\begin{array}{c} a(t)\\{}\\ b(t)\end{array}\right)\,,
\end{equation}
with eigenvalues $\pm\omega_p |c_n|/2$, leading to growth exponent
\begin{equation}\label{growth4}
    \omega''_n = \frac{\omega_p |c_n|}{2} 
\end{equation}
for the instability at modulation frequency $\Omega = 2\omega_p/n$. This result is linear in $c_n$, and therefore in $\deln$ by construction. Thus, any such parametric resonance is one of the set of {\em dominant} resonances for the modulation profile $\deln(t)$, analogous to the one associated with \eqref{growth1} in the case of piecewise constant modulation (see Sect. \ref{Sec_piecewise}). 

We comment that in the averaging procedure leading to \eqref{hamiltonian3} we have obviously lost all information about the oscillatory behavior of the amplitudes $a(t)$ and $b(t)$, and in particular, of the real part $\omega'$ of the Floquet frequency corresponding to this resonance. Indeed, the (approximate) amplitudes $a(t)$ and $b(t)$ resulting from \eqref{hamiltonian3} are not oscillatory and have purely real exponential behavior $e^{\pm \omega'' t}$. The only oscillatory part of the full solution for $\Psi (t) = (\psi_1(t), \psi_2(t))^T$ comes from the phases $e^{\pm i\omega_p t}$ in \eqref{spinor}, which are independent of $\deln(t)$ and therefore of the modulation frequency $\Omega$. Indeed, the approximate growing (unstable) solution arising from \eqref{hamiltonian3} and \eqref{growth4} is $(a(t), b(t))^T \simeq (1,-i)^T e^{\omega'' t}$, leading to $(\psi_1,\psi_2)^T = (\omega_p f(t), \dot f(t))\simeq  (\cos \omega_p t, -\sin \omega_p t) e^{\omega'' t}$ (where a term of order $\omega''/\omega_p\propto |c_n|$ was neglected in the component $\psi_2 = \dot f$). Thus, the cosine factor in $f(t)$ oscillates periodically at frequency $\omega_p = n\Omega/2 \neq \Omega$, and is therefore not even a proper Floquet solution (unless $n=2$ accidentally). Therefore, the averaged `hamiltonian' \eqref{hamiltonian3} should only be used to determine the exponential envelope of the Floquet eigensolutions. 

In Sect. \ref{Sec_comparison} we compare the result predicted by Eq.~$\eqref{growth4}$ with the exact result for different modulation schemes.

\subsection{Optimal modulation profile}

For a weak modulation, the gain rate for the $n$-th resonance is given by \eqref{growth4} with
\begin{eqnarray}
c_n  = \frac{1}{T}\int\limits_0^T { \deln \left( t \right)e^{ - in\Omega t} } dt.
\end{eqnarray}
In the following, we demonstrate that if the modulation depth is bounded as $\left| \deln (t) \right| < \delnM$, then the optimal modulation profile that maximizes the gain rate in the weak modulation regime corresponds to a piecewise constant modulation. For simplicity, we focus the discussion on the $n=1$ case, though the derivation can be easily generalized to arbitrary $n$.

To begin with, we note that the coefficient $c_1$ satisfies 
\begin{subequations} 
\begin{eqnarray}  
\left| {c_1 } \right| = \sqrt {a_1^2  + b_1^2 } \quad ,{\rm   with} \\ 
 a_1  = \frac{1}{T}\int\limits_0^T {\deln \left( t \right)\cos \left( {\Omega t} \right)} dt \\ 
 b_1  = \frac{1}{T}\int\limits_0^T {\deln \left( t \right)\sin \left( {\Omega t} \right)} dt
\end{eqnarray}
\end{subequations}
Based on these identities, one can also write:
\begin{eqnarray}   \label{optaux} \left| {c_1 } \right| &=&  a_1 \cos \phi  + b_1 \sin \phi  \nonumber \\ 
  &=&  \frac{1}{T}\int\limits_0^T {\delta _{\bmod } \left( t \right)\cos \left( {\Omega t - \phi } \right)} dt 
\end{eqnarray}
with $
\left( {\cos \phi ,\sin \phi } \right) = \left( {a_1 ,b_1 } \right)/\sqrt {a_1^2  + b_1^2 }$. From the second identity of Eq.~\eqref{optaux}, we obtain the following bound for $\left| {c_1 } \right|$ in terms of the peak of the modulation depth:
\begin{eqnarray}  
\left| {c_1 } \right| \le \delta _{{\rm{mM}}} \frac{1}{T}\int\limits_0^T {\left| {\cos \left( {\Omega t-\phi} \right)} \right|} dt = \delta _{{\rm{mM}}} \frac{2}{\pi }.
\end{eqnarray}  
Hence, in the weak modulation regime, the gain rate of the main resonance is bounded as:
\begin{eqnarray}  
\omega ''_1  \le \frac{{\omega _p }}{\pi }\delta _{{\rm{mM}}}.
\end{eqnarray} 
It is shown in Sect. \ref{Sec_piecewise} that the bound is saturated for a piecewise constant modulation with 50\% duty cycle (see Eq.~\eqref{growth1}). Thus, it follows that the piecewise constant modulation provides the optimal modulation profile to maximize the gain rate.

\section{Piecewise Constant Modulation}
\label{Sec_piecewise}

Here, we study in detail the piecewise constant modulation scheme and its impact on the longitudinal plasmons. In particular, we derive the expressions for the growth rate coefficients associated with resonances of odd and even indices. We also solve this model in the limit in which the piecewise modulation becomes a time-periodic Dirac comb of $\delta$-impulses. 

\subsection{Modulation Period with Two Different Amplitude Values}

In the simplest form of piecewise constant periodic modulation, 
$\deln (t)$ assumes one constant value $\delta_{\rm mod1}$ during the first part $0\leq t < \tau$ of the modulation period and another constant value $\delta_{\rm mod2}$ during its remaining part $\tau \leq t <T$ \cite{Brillouin, vdPS}. 

The common practice to solving \eqref{divergenceg} with periodic modulation is to analyze the initial value problem, and compute the two independent fundamental solutions $c(t)$ and $s(t)$ with initial conditions $c(0)=1;\,\dot c(0)=0$ and $s(0)=0;\,,\dot s(0)=1$ (namely, the cosine- and sine-like solutions). These solutions can be easily computed explicitly in our case of piecewise constant modulation. The transfer matrix $\M{T}$ is then constructed from the values of these fundamental solutions and their derivatives at $t=T$\,\cite{Hill}. Here we shall follow an alternative method\cite{Serra2, Prud, Ioannis} (familiar from the theory of Lyapunov stability) to computing $\M{T}$, but in a basis different from that of $\{ c(t), s(t)\}$. (The eigenvalues $\Lambda_{1,2}$ of $~\M{T}$ are of course basis independent.) 

Specifically, we follow the general Hamiltonian-type approach outlined in Sect. \ref{Sec_propagator}.
For constant $\deln$, as in each part of the modulation period in the present example, one can exponentiate $\M{L}$ explicitly and find 
\begin{equation}
    \label{exponential}
    \M{M}(\Theta;t)  = e^{\M{L}t}  = \left(\begin{array}{cc} \cos \Theta t & \frac{\omega_p}{\Theta}\sin \Theta t\\{}\\ -\frac{\Theta}{\omega_p} \sin \Theta t & \cos \Theta t\end{array}\right)
\end{equation}
with 
$\Theta = \omega_p \sqrt{1+ \deln}$\,. 
In the main text we focus on the modulation 
\begin{equation}\label{piecewise}
\delS\left( t \right) = \delnM\, {\mathop{\rm sgn}} \left( {\sin \left( {\Omega t} \right)} \right)
\end{equation} 
which flips sign at the middle of the modulation period. In this case the transfer matrix for Eq.~\eqref{divergenceg} is readily found as $\M{T}_g = \M{U}(T,0) = \M{M}(\Theta_-;T/2) \M{M}(\Theta_+;T/2)$, where $\Theta_{\pm} = \omega_p \sqrt{1-\frac{\nu^2}{4\omega_p^2} \pm \delnM}$. For simplicity, here we assume that the dissipation rate is time independent.

From the relation $f=e^{-\nu t/2}g$ we have $(\omega_p f(t),\dot f(t))^T =e^{-\nu t/2} \M{K} (\omega_p g(t), \dot g(t))^T$ with $\M{K} = \left(\begin{array}{cc} 1 & 0\\ -\nu/2\omega_p  & 1\end{array}\right)$, and therefore the transfer matrix for Eq. \eqref{divergence1} is $\M{T}_f = e^{-\nu T/2} \M{K} \M{T}_g\M{K}^{-1}.$ Thus, 
\begin{eqnarray}\label{dispersion2}
    &&\frac{1}{2}{\rm Tr}\,\M{T}_f =\frac{1}{2}(\Lambda_1 + \Lambda_2) = \frac{1}{2}e^{-\nu T/2} {\rm Tr}\,\M{T}_g \nonumber\\
    &&=e^{-\nu T/2} \left[\cos\left(\frac{\pi\Theta_+}{\Omega}\right)\cos\left(\frac{\pi\Theta_-}{\Omega}\right)\right.\nonumber\\{}\nonumber\\ && \left. - \frac{1}{2}\left(\frac{\Theta_+}{\Theta_-}+\frac{\Theta_-}{\Theta_+}\right)\sin\left(\frac{\pi\Theta_+}{\Omega}\right)\sin\left(\frac{\pi\Theta_-}{\Omega}\right)\right]
\end{eqnarray}
which coincides with the known result \cite{vdPS, Brillouin} in the absence of dissipation $\nu=0$. (See also chapter 8 of\, \cite{Hill} and note the obvious typo in Eq.(8.5) therein.) 

For simplicity, from now on we shall focus on the non-dissipative case $\nu=0$, and rewrite \eqref{dispersion2} more neatly as 
\begin{eqnarray}\label{dispersion}
    &\frac{1}{2}&{\rm Tr}\,\M{T}_f = \cos\left(\frac{\pi (\Theta_+ + \Theta_-)}{\Omega}\right)\nonumber\\&-& \frac{(\Theta_+ - \Theta_-)^2}{4\Theta_+\Theta_-}\cdot\left[\cos\left(\frac{\pi (\Theta_+ - \Theta_-)}{\Omega}\right)\right.\nonumber\\
    &&\left. - \cos\left(\frac{\pi (\Theta_+ + \Theta_-)}{\Omega}\right)\right]\,.
\end{eqnarray}
In regions of stability, where $\Lambda_2^* = \Lambda_1 = e^{-i\omega T}$ (with real $\omega$), we thus have $\frac{1}{2}{\rm Tr}\,\M{T}_f = \cos \omega T$, so that $|{\rm Tr}\,\M{T}_f/2| < 1$. In contrast, in the unstable regime we can always choose $\Lambda_1=1/\Lambda_2 = \pm e^{\omega'' T}$ with $\omega'' >0$. Thus, $\omega = i\omega ''$ for positive $\Lambda_{1,2}$, or $\omega =\pi/T + i\omega ''$ for negative $\Lambda_{1,2}$ (where $\omega' = \pi/T$ is restricted to the first Brillouin zone). Therefore, in the unstable regime we have $\frac{1}{2}{\rm Tr}\,\M{T}_f = \pm\cosh (\omega''T)$ so that $|{\rm Tr}\,\M{T}_f/2|> 1$. The boundaries separating stable and unstable regions are therefore given by the curves where  ${\rm Tr}\,\M{T}_f/2 =\pm 1$. This leads, like in the case of Mathieu's equation, to a rich chart of stability and instability regions in the plane of parameters $\Omega/\omega_p$ and $\delnM$\cite{Brillouin, vdPS}.  

Let us now investigate the stability of this system perturbatively for weak modulation amplitude $\delnM$. In this case, the factor $(\Theta_+-\Theta_-)^2/\Theta_+\Theta_-$ in the second term in \eqref{dispersion} is clearly of order $\delnM^2$. Thus, the modulation frequencies which induce instabilities at infinitesimal $\delnM\rightarrow 0^+$ are determined in this limit by the leading term $\cos (\pi(\Theta_+ + \Theta_-)/\Omega)\simeq \cos (2\pi\omega_p/\Omega)$ in \eqref{dispersion} tending to $\pm 1$. The first few terms of the expansion of \eqref{dispersion} around $\delnM=0$ are
\begin{eqnarray}\label{dispersion1}
   &&\frac{1}{2}{\rm Tr}\,\M{T}_f = \cos\left(\frac{2\pi\omega_p}{\Omega}\right) + \nonumber\\ &&\frac{1}{2}\left[\frac{\pi\omega_p}{2\Omega}\sin \left(\frac{2\pi\omega_p}{\Omega}\right) - \sin^2\left(\frac{\pi\omega_p}{\Omega}\right)\right]\delnM^2 +\nonumber\\ && \frac{1}{32}\left[4\left(\frac{\pi \omega_p}{\Omega}\right)^2 -6 +\left(6 - \left(\frac{\pi \omega_p}{\Omega}\right)^2\right)\cos\left(\frac{2\pi \omega_p}{\Omega}\right) +\right.\nonumber\\ && \left. \frac{9}{2}\frac{\pi \omega_p}{\Omega}\sin\left(\frac{2\pi \omega_p}{\Omega}\right)\right]\delnM^4 + {\cal O}(\delnM^6)\,.\nonumber\\
\end{eqnarray}
The instability near $\cos (2\pi\omega_p/\Omega) = -1$ occurs around modulation frequencies 
$\Omega = 2\pi/T= 2\omega_p/(2n+1)$ with integer $n$. (In the expansion of \eqref{dispersion} around $\delnM=0$ leading to \eqref{dispersion1}, we tacitly assumed $T$ was bounded, which means that the integer $n$, indexing the corresponding parametric resonance, is assumed to be bounded as well.) Thus, let us substitute $\Omega = 2\omega_p(1 + \Delta)/(2n+1)$ in \eqref{dispersion1}, with $\Delta$ a small detuning parameter. We find
$1/2{\rm Tr}\,\M{T}_f \equiv \cos\left((\omega'+i\omega'') T\right)=   -1 -\left[\delnM^2 -(2n+1)^2\pi^2\Delta^2\right]/2 + {\cal O}(\delnM^2\Delta, \Delta^3)\simeq \cos\left(i\sqrt{\delnM^2-(2n+1)^2\pi^2\Delta^2}\pm \pi\right)$. Therefore, up to the indicated accuracy, $\omega' = \pm\pi/T = \pm\omega_p(1+\Delta)/(2n+1)$ and 
\begin{equation}\label{growth1}
\omega'' = \omega_p\sqrt{\frac{\delnM^2}{((2n+1)\pi)^2} - \Delta^2}\,. 
\end{equation}
Thus, maximal instability (the center of the resonance) occurs at $\Delta = 0$, where $\omega''_{\rm Res} = \omega_p \delnM/(2n+1)\pi$ is linear in $\delnM$, and is therefore of leading order in perturbation theory. Furthermore, similarly to the corresponding result for the Mathieu case discussed in the main text, detuning the modulation frequency away from the resonance 
leads to a reduction of the growth rate $\omega''$. The instability persists for $|\Delta| < \delnM / [(2n+1)\pi]$

This is not the case for the parametric resonances around the other instability borderline at $\cos (2\pi\omega_p/\Omega) = +1$, which occur in the vicinity of modulation frequencies $\Omega = 2\omega_p/2n = \omega_p/n  $ with integer $n$. By substituting $\Omega = \omega_p(1 + \Delta)/n$ in \eqref{dispersion1} and expanding the resulting expression in powers of $\Delta$, we find that the associated resonance lies in the parametric regime where $\Delta$ scales like $\delnM^2$ so that $1/2{\rm Tr}\,\M{T}_f \equiv \cos\left((\omega'+i\omega'') T\right)= 1 +(1/2)(2\pi n)^2[\delnM^4/16 - (\Delta + \delnM^2/8)^2] + {\cal O}(\Delta^3, \delnM^2\Delta^2, \delnM^4\Delta, \delnM^6)\simeq \cosh\left[(2\pi n)\sqrt{\delnM^4/16-(\Delta + \delnM^2/8)^2}\right]$. Thus, up to the indicated accuracy 
\begin{equation}\label{growth3}
\omega'' = \omega_p\sqrt{\left(\frac{\delnM^2}{4}\right)^2 - \left(\Delta + \frac{\delnM^2}{8}\right)^2}
\end{equation}
(and of course $\omega'=0$). Thus, in contrast with \eqref{growth1}, these resonances (that is, maximal instability) are not centered at modulation frequency $\Omega = \omega_p/n$ where $\cos (\omega T) = 1$ reaches the upper border of the stability region, but rather at a slightly smaller and $\delnM$-dependent modulation frequency $\Omega_{\rm Res} = (1-\delnM^2/8)\omega_p/n$. Moreover, notwithstanding the $n$-dependence of the location resonances of this type, \eqref{growth3} is {\em completely independent of} $n$, in contrast with \eqref{growth1}, with a {\em common} maximal growth exponent $\omega''_{\rm Res} = \omega_p \delnM^2/4$, which is quadratic in $\delnM$, and is therefore of higher order in perturbation theory.  

\subsection{Effect of Dissipation on Longitudinal Plasmons - Numerical Results}
\begin{figure}[h!]
\includegraphics[scale=1.]{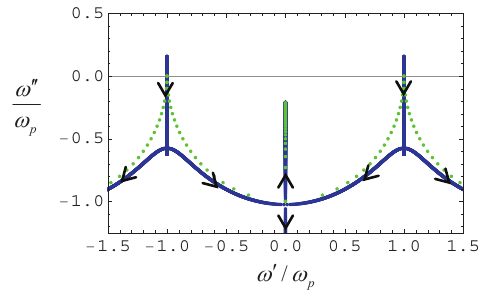}
\caption{ Locus of $\omega' + i\omega''$ in the complex plane as a function of the damping strength $\nu$ at modulation frequency $\Omega = 2\omega_p$ for the piecewise constant modulation \eqref{piecewise}. The arrows indicate the direction of increasing $\nu$. i) $\delnM=0.5$ (solid blue lines), ii) $\delnM=0^+$ (green dots). The horizontal grid line (in gray) separates the stable and unstable regions.
}
\label{Fig-Dissipation}
\end{figure}

The discussion in the main text is focused on lossless systems, $\nu=0$. For completeness, in Fig.~\ref{Fig-Dissipation} we demonstrate schematically the impact of the damping coefficient $\nu$ on longitudinal plasmons (for the modulation profile in \eqref{piecewise}). 

This figure shows the locus of the eigenfrequencies $\omega' + i \omega''$ in the complex plane, for values of the loss parameter $\nu$ in the range $0<\nu < 5 \omega_p$. The direction of increasing loss is indicated by arrows. The modulation frequency is $\Omega = 2 \omega_p$ and $\delnM = 0.5$ (strong modulation). At $\nu=0$, one of the eigenfrequencies resides in the upper half of the frequency plane, represented by the upper endpoints of the curves (note that due to the  periodicity of the band diagram in $\omega'$ the two endpoints represent the same eigenfrequency). Its counterpart (not singled out in the figure), is located symmetrically across the $\omega''=0$ horizontal axis. As the dissipation parameter increases, both modes descend towards the lower-half frequency plane, eventually converging in a bifurcation. The eigenfrequency of the mode experiencing gain intersects the $\omega''=0$ line when $\nu$ is approximately $0.32 \omega_p$, highlighting a critical transition point influenced by dissipation. For comparison, the green points represent a similar study for the case when the modulation strength is vanishingly small ($\delnM \approx 0$). As expected, in this case the spectrum lies completely in the lower-half frequency plane.

\subsection{Dirac Comb of $\delta$-Impulses}
Going back to the more general piecewise constant profile (with uneven durations of the two constant values of the modulation amplitude $\delS$), an interesting limit is obtained when, for example, we let $\tau\rightarrow 0^+, \delta_{m1}\propto 1/\tau$ and $\delta_{m2}=0$. Here, $\tau$ stands for the duration of the first cycle of the modulation period. In this limit we obtain a Dirac comb of $\delta$-impulses, namely,
\begin{equation}\label{kicks}
\delS(t) =  \alpha\sum_{j=1}^\infty \delta(t-jT)
\end{equation}
with time-independent parameters $\alpha$ and $T$. With this type of modulation, Eq. \eqref{divergence1} is conveniently solved by computing explicitly the transfer matrix $\M{T}$ (in the `scattering basis'), as we now explain. Between impulses, say for $jT< t<(j+1)T$, $f(t)$ evolves as a linear combination $A_je^{-i\omega_+(t-jT)} + B_je^{-i\omega_-(t-jT)}$ of the two frequencies $\omega_\pm = -i\nu/2 \pm\omega'_p$ of the time-independent problem, with $\omega'_p = \sqrt{\omega_{p}^2 - \nu^2/4}$, under the further assumption that 
$\omega'_p$ is real-valued, namely, that the system is not over-damped. The solution $f(t)$ is continuous throughout the impulse at $t=(j+1)T$, while $\partial_t f$ suffers a jump discontinuity: ${\partial _t}{f_ + } - {\partial _t}{f_ - } =- \alpha \omega _p^2 f$ (with obvious notations).

These properties uniquely determine the amplitudes $A_{j+1}$ and $B_{j+1}$ right after a given impulse as $(A_{j+1},B_{j+1})^T = \M{T}(A_j,B_j)^T$ where\cite{6} 
\begin{equation}\label{transfer}
\M{T}  = e^{-\frac{\nu T}{2}}\left(\!\!\begin{array}{cc}1-iu & -iu\\{}\\ iu & 1+iu\end{array}\right) \left(\begin{array}{cc} e^{-i\theta} &  0\\{}\\0 & e^{i\theta} \end{array}\right)
\end{equation}
with 
$u = \frac{\alpha \omega_p ^2}{2\omega'_p}$ and $\theta=\omega'_p T$.
The eigenvalues of $\M{T}$ control stability of the system, namely, whether the amplitudes grow (regions of instability in parameter space) or remain bounded (regions of stability). For simplicity, let us analyze the stability properties of the PLTC in the absence of damping ($\nu=0$), where $u=\frac{\alpha\omega_p}{2}$. 

In this case $\det \M{T}=1$ and the two mutually reciprocal eigenvalues of $\M{T}$ are
$\Lambda_\pm = 
{\rm{Tr}}\, {\M{T}}/2
\pm \sqrt{({\rm{Tr}}\,{\M{T}}/2)^2-1}$, where ${\rm{Tr}}\,{\M{T}} /2 = {\mathop{\rm Re}\nolimits}\,T_{11} = \cos \theta  - u\sin \theta$.

Thus, if $|{\rm Re}T_{11}|>1$, both eigenvalues are real, with either $|\Lambda_+|$ or $|\Lambda_-| >1$, resulting in the growth of $f(t)$ after each kick. This is the region of instability. If, on the other hand, $|{\rm Re}T_{11}|<1$, then $\Lambda_-=\Lambda_+^*$ so that $|\Lambda_\pm|=1$ and $f(t)$ remains bounded as function of time. This is the region of stability. Clearly, the boundaries separating regions of stability and instability are determined by  
$ \cos\theta -u\sin\theta  = \pm1$. Such points in the parameter space of $\M{T}$ are exceptional points, where the matrix is non-diagonalizable, possessing only a single eigenvector.
\begin{figure}[h!]
\includegraphics[scale=1.]{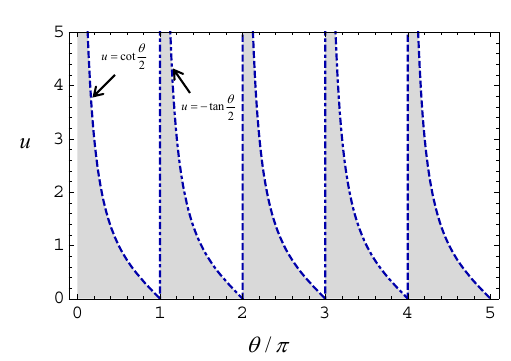}
\caption{ Stability region of the $u-\theta$ plane for the Dirac impulse model (\ref{kicks}). The zone shaded in gray is the stability region in parameter space. The boundary curves are determined by $u = \cot \theta/2 $ (dashed curves) and $u = -\tan \theta/2$ (dot dashed curves).
}
\label{FigKicks}
\end{figure}
Straightforward calculation shows that at the instability threshold $u$ and $\theta$ are related either by $u = \cot \theta/2 $ (when ${\rm{Tr}}\,{\M{T}} /2 = -1$) or $u = -\tan \theta/2$ (when ${\rm{Tr}}\,{\M{T}} /2 = 1$). Let us denote the border stability lines of the first type by $\theta_-(u)$, and the border stability lines of the second type by $\theta_+(u)$. As illustrated in Fig.~\ref{FigKicks}, the stability regions are the areas lying between these curves and the horizontal axis in the $u-\theta$ plane. The curves $\theta_-(u)$ terminate on the horizontal axis of Fig.~\ref{FigKicks} at points where $\theta/\pi = $ {\em an odd integer}, whereas the curves $\theta_+(u)$ terminate there at points where $\theta/\pi = $ {\em an even integer}. Let us consider now narrow strips in the unstable region immediately to the right of the border lines. That is, for a given $u$, we set $\theta = \theta_{\pm}(u) + \Delta$ with $0<u\Delta\ll 1$. By expressing $\sin\theta_{\pm}(u)$ and $\cos\theta_{\pm}(u)$ in terms of $u$, it is straightforward to show in these strips that $\cos (\omega T) = {\rm{Tr}}\,{\M{T}} /2 =\pm(\cos\Delta + u\sin\Delta)\simeq \pm \cosh (\sqrt{2u\Delta})$. Thus, immediately on the right of both lines $\theta_{\pm}(u)$ the instability growth rate is $\omega'' = \sqrt{2u\Delta}/T$, but such generic instability is non-resonant (in the sense that it does not have a local maximum as function of $\Delta$). Resonances naturally appear near the horizontal axis in Fig.~\ref{FigKicks}, where the 'tongues' of instability terminate, that is where also $\theta/\pi=n$ an integer. (We remind the reader that odd integers correspond to the $\theta_-(u)$ lines, and even integers to the $\theta_+(u)$ lines.) Thus, for $u\sim \Delta\ll 1$ we obtain $\cos (\omega T) =\pm(\cos\Delta + u\sin\Delta)\simeq (-1)^n\cosh (\sqrt{u^2-(\Delta-u)^2}) + {\cal O}(\Delta^4, u\Delta^3)$. Therefore, in the {\em vicinity} of a given $\theta \simeq n\pi$ in the $u-\theta$ plane, the resonance is centered at $\Delta =u$ (that is, at $\theta_{\rm Res} = \theta_\pm(u) + u$), with maximal growth rate 
\begin{equation}\label{growth-Dirac}
\omega_n'' = \frac{u}{T} = \frac{\Omega u}{2\pi}\,,
\end{equation}
{\em independently} of $n$. The real part of the Floquet frequency depends on the the parity of $n$. Thus, $\omega'=0$ for resonances corresponding to even $n$, while $\omega' = \pm\pi/T = \pm \Omega/2$ (in the first Brillouin zone) for resonances corresponding to odd $n$. 

Finally, as can be seen in Fig.~\ref{FigKicks}, the system is always unstable in the vicinity of $\theta_n = n \pi+0^-$ (with integer $n$), which corresponds to modulation frequency $\Omega_n = 2 \omega_p / n$. Furthermore, as $u\rightarrow 0+$ (no impulses) the gaps disappear, rendering the system always stable, while in the opposite limit $u\rightarrow +\infty$ the stability regions shrink to the points $\theta_n = n\pi + 0^+$. 

\subsection{Comparison with Eq. ~\eqref{growth4}}
\label{Sec_comparison} 

Next, we compare the result predicted by Eq.~$\eqref{growth4}$ with the exact result for the different modulation schemes.

For example, in the Mathieu case (see the main text),  $c_l = (\delnM/2)\delta_{l,1}$, so, according to Eq.~$\eqref{growth4}$, there is only one dominant resonance which occurs for $n=1$ at $\Omega=2\omega_p$ and with the known growth exponent $\omega'' =\omega_p\delnM/4$. 

For the piecewise constant modulation \eqref{piecewise}, the Fourier modes are $c_n=0$ for even $n$ and $c_n = 2\delnM/\pi i n$ for odd values of $n$. Thus, there is an {\em infinite set of dominant resonances} at modulation frequencies $\Omega = 2\omega_p/n$ with $n$ an odd number, with corresponding growth exponents $\omega'' = \omega_p \delnM/\pi n$, which agrees with the result \eqref{growth1} obtained directly from the dispersion relation \eqref{dispersion}. There are no {\em dominant} resonances corresponding to $n$ even simply because there are no Fourier modes of even order for the modulation \eqref{piecewise}, as we discovered by direct analysis of \eqref{dispersion1}. 

Finally, for the Dirac-comb impulse modulation \eqref{kicks}, the Fourier modes are $c_n = \frac{\alpha}{T} = \frac{2u}{\omega_p T}$, independently of $n$, and indeed, we see from \eqref{growth-Dirac} and \eqref{growth4} that 
$\omega'' = \frac{u}{T} = \frac{\omega_p |c_n|}{2}$. 

In order to determine higher order resonances of the modulated system, that is, resonances with growth exponents $\omega''$ which depend on higher powers of the small amplitude $\deln$, as those mentioned in \cite{LL} for the Mathieu equation, or those corresponding to \eqref{growth3} for the piecewise constant case, one should analyze higher orders in the time-dependent perturbative expansion of the solution of \eqref{hamiltonian2}.

\section{Numerical Results for Resonance Growth Rates for Transverse Modulated Waves}
In this section, we neglect dissipation and assume that $\delta_{\varepsilon}=0$. Thus, we only consider the effects of the modulation of the electron effective mass by the pump signal.
From the analysis in Sect.~\ref{Sec_RelationLongTrans}, for the non-dissipative case, the discussion of instabilities in longitudinal modes can be transferred, mutatis mutandis, to transverse modes. 

The most significant instability is linked to the interband transition at modulation frequency $\Omega_T(k) = 2\omega_T(k) = 2\sqrt{\omega_p^2 + c^2k^2/\varepsilon_0}$, as indicated in Fig.~1b in the main text. As in the case of longitudinal plasmons, we interpret this instability as arising from the interaction between negative and positive frequency branches $\pm\omega_T(k)$. However, for transverse plasmons, it is $k$-dependent, in contrast with the longitudinal case. Consequently, near the main resonance, the parametric gain is predominantly governed by longitudinal plasmons, which, as discussed in the main text, experience a collective, $k$ independent leading resonance at $\Omega =  2\omega_p$. It is this collective resonance that determines the optimal path for gain extraction from the external driving. 

\begin{figure}[th!]
\includegraphics[scale=1.]{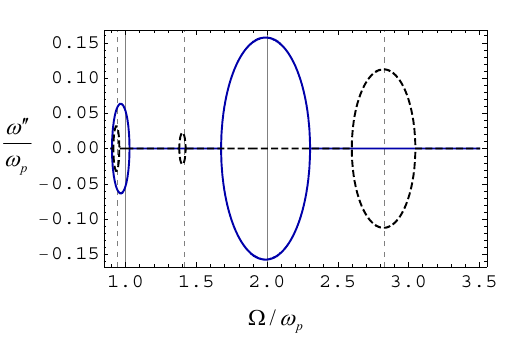}
\caption{
Amplification rate as a function of the modulation frequency for the same system as in Fig.~2b in the main text. Solid blue lines: longitudinal plasmons and transverse plasmons with $k=0$. Dashed black lines: Transverse plasmons with $k = \sqrt{ \varepsilon_0 }\omega_p/c$, that is, $\omega_T(k) = \sqrt{2}\omega_p$. The two solid vertical gridlines coincide with their counterparts at $\Omega/\omega_p = 1,2$ in Fig. 2 in the main text. The dashed vertical lines indicate modulation frequencies $\Omega = 2\omega_T/m = 2\sqrt{2}\omega_p/m$ for $m=1,2,3$.
}
\label{FigTransverse}
\end{figure}

For completeness, we show in Fig.~\ref{FigTransverse} the gain rate of transverse plasmons as a function of the modulation frequency in the case of piecewise modulation, for a particular value of $k$. The system parameters are as in Fig. 2b in the main text, namely, strong modulation amplitude $\delnM=0.5$. Note that at such strong modulation amplitudes, the  locations of longitudinal resonances shift from their corresponding weak amplitude locations around $\Omega = 2\omega_p/m$ with integer $m$. However, as is evident from Fig. 2b, these shifts are bounded, and at least for the first few (longitudinal and transverse) resonances we discuss specifically in this section, there is no risk of confusion among resonances by referring to the corresponding locations at weak amplitude modulation. Having said that, we see in Fig.~\ref{FigTransverse} that the gain rate for $k=0$ coincides with the gain rate of the longitudinal plasmons (solid blue curves). (Shown in Fig.~\ref{FigTransverse} are the two resonances corresponding to values $m=1,2$ of Fig. 2b at $\Omega=2\omega_p$ and $\omega_p$, respectively.) For nonzero $k$ the gain peaks are blue-shifted (dashed black curves) to $\Omega = 2\omega_T/m > 2\omega_p/m$ with integer $m$. 

Using the results in Sect.~\ref{Sec_RelationLongTrans} (with $\delta_{\varepsilon}(t)=0$), we can relate the transverse resonances to those arising in the longitudinal case.
Specifically, the growth rates for corresponding resonances are linked as $
\omega ''_{n,{\rm{trans}}}  = \frac{{\omega _p }}{{\omega _T }}\omega ''_{n,{\rm{long}}} 
$.
Thus, for nonzero $k$, the growth rate for the transverse waves is smaller than for the corresponding longitudinal waves.

For the particular chosen value of $k$ in Fig.~\ref{FigTransverse} we have $\omega_T = \sqrt{2}\omega_p$. As expected, the displayed dashed curves, indicating transversal resonances are centered near the dashed vertical lines at $\Omega/\omega_p = 2\sqrt{2}, \sqrt{2}$ and $2\sqrt{2}/3$, corresponding to the predicted (weak amplitude locations) at $2\omega_T/m$ with $m = 1,2,3$. As is also evident from Fig.~\ref{FigTransverse}, the peak gain for the transverse resonances at $k\neq 0$ are smaller compared to their counterparts in Fig. 2b, which is consistent with the effective reduction of the effective modulation amplitude, as discussed above.